\documentclass[final]{svjour3}
\usepackage{graphicx}
\usepackage{rotating}
\usepackage{amssymb}
\usepackage{mathptmx}
\usepackage{cite}
\usepackage{subfig}
\usepackage{diagbox}
\usepackage{colortbl}
\makeatletter
\journalname{Journal of Low Temperature Physics}
\begin{document}

\newcommand{\hdblarrow}{H\makebox[0.9ex][l]{$\downdownarrows$}-}
\title{Modelling signal oscillations arising from electro-thermal coupling and stray capacitance in semiconducting bolometer impulse response}
\titlerunning{Demonstrating signal oscillations due to electro-thermal feedback and stray capacitance}

\author{S. L. Stever\textsuperscript{{\normalfont \textit{a,b}}} \and F. Couchot\textsuperscript{{\normalfont \textit{c}}}}

\institute{
\textsuperscript{\textit a}Okayama University,  3-1-1, Tsushimanaka, Kita-ku, Okayama City, Okayama, 700-8530, Japan,
\textsuperscript{\textit b}Kavli IPMU (WPI), UTIAS, The University of Tokyo, Kashiwa, Chiba 277-8583, Japan, 
\textsuperscript{\textit c}Université Paris-Saclay, Laboratoire de Physique des 2 Infinis Irène Joliot Curie, Orsay, 91400, France
} 

\maketitle

\begin{abstract}
Electro-thermal coupling in semiconductor bolometers is known to create nonlinearities in transient detector response, particularly when such detectors are biased outside of their ideal regions (i.e. past the turnover point in their IV curves). This effect is further compounded in the case where a stray capacitance in the bias circuit is present, for example in long cryogenic cabling. We present a physical model of the influence of such electro-thermal coupling and stray capacitance in a composite NTD germanium bolometer, in which previous experimental data at high $V_{\rm bias}$ resulted in oscillations of the impulse response of the detector to irradiation by alpha particles. The model reproduces the transient oscillations seen in the experimental data, depending both on electro-thermal coupling and stray capacitance. This is intended as an experimental and simulated example of such oscillations, demonstrated for the specific case of this bolometric detector.
\end{abstract}

\vspace{-6mm}
\section{Introduction}
\vspace{-3mm}
Characterising the impulse response of bolometric detectors depends not only on the nature of the impulse (e.g. ionising radiation or photons) but also on the steady-state properties of the bias circuit and the level of the detector bias. In the case of semiconductor bolometers, the impulse response is partially defined by current-voltage characteristics (i.e. the load curve) as well as the resistance as a function of temperature ($R(T)$). In particular, if a bolometer is operated past the turnover point in its dark load curve, certain electro-thermal effects become dominant, affecting the impulse response with nonlinear effects. For instance, it is known that stray capacitance in bolometer electrical cabling can induce signal oscillations in the bolometric response function in this regime~\cite{lindeman}. 

This manuscript is intended as one clear example of this effect. We will describe experiments performed in the characterisation of the behaviour of a cryogenically-cooled composite semiconductor NTD Ge bolometer irradiated by $\alpha$ particles at 100 mK at an unstable working point, in which oscillatory features were noted. In order to confirm the nature of the oscillations as being due to electro-thermal effects, we have modelled the bolometer circuit, accounting for the presence of electro-thermal feedback in the detector and stray capacitance in the bias circuit. This serves only as one such example of modelling electrothermal oscillations in the wider body of work; another example can be found in e.g. Nutini et al. \cite{nutini}.

\vspace{-6mm}
\section{Thermal and electrical description}\label{sec:TES}
\vspace{-3mm}

Bolometers are classically described (for instance in the Mather model~\cite{mather}) as a radiation-absorbing thermistor with temperature $T_{B}$, a resistance $R_{\rm bolo}$, and a heat capacity $C$ coupled to a thermal bath held at temperature $T_{0}$ through a thermal link with a thermal conductance $G$. Incident optical radiation (power) $P_{\rm opt}$ on the thermistor changes its temperature. Using a bias circuit, a resistor in series with a voltage source generates a current $I_{\rm bias}$ through the thermistor, resulting in a potential difference $V = I_{\rm bias}R_{\rm bolo}$. The power dissipated into the thermistor by the biasing mechanism, $P_{\rm Joule}$, is then combined with the incident optical power to create a total power dissipated into the thermistor of $W = P_{\rm Joule} + P_{\rm opt}$. The time taken to diffuse heat from this total power in the thermistor to the thermal bath, in the absence of electro-thermal feedback, is determined by its physical thermal time constant $\tau = C / G$.

Further expansion on Mather's theorem of non-equillibrium bolometers was examined in the Griffin and Holland model~\cite{griffin} and further by Sudiwala et al.~\cite{rashmi}, leading to the more realistic case of thermal conductance $G$ scaling as a function of temperature. This can be expressed in terms of a power law where $G_{D}(T_{\rm bolo},T_{0}) \times (T_{\rm bolo} - T_{0}) = g[T_{\rm bolo}^{\beta+1} - T_{0}^{\beta+1}]$ and $G_{0}$ is the baseline thermal conductance when $T_{\rm bolo} = T_{0}$. Therefore, in a steady state, the equillibrium point is reached when the power dissipated in the thermistor achieves the condition that $W = G_{D}(T_{\rm bolo} - T_{0})$:

%, which assume that the thermal link itself also has a temperature-variable thermal conductance $k$ defined by a power law, leading to a new definition for the equillibrium state as:

\begin{equation}
    W = \frac{A}{L}\frac{k_{0}T_{0}}{\beta+1}\left(\frac{T_{\rm bolo}}{T_{0}}^{\beta +1} - 1\right)
        \label{eq:W_Griffin}
\end{equation}

\noindent where $A$ is the cross-sectional area of the thermal link, $L$ is its length, and $k_{0}$ is the thermal conductance at $T_{0}$.

%%At the working point, a $\Delta T$ induced by input power on the bolometer 

\vspace{-7mm}
\section{Experimental background}
The bolometer used in the experiment (``Bolo 184'') is a composite semiconductor bolometer, containing a disc-shaped diamond absorber with a sputtered back-layer of bismuth, an NTD germanium sensor, and a sapphire slab for mechanical support. The bolometer layers are held together using Devcon epoxy, and the ends of the long, thin germanium and sapphire legs are coupled to the thermal bath using wound copper wire. The dimensions of the various components of the bolometer are given in Table~\ref{tbl:Bolo184}.

\begin{table}[t]
\begin{tabular}{llll}
\textbf{Component}                        & \textbf{Width}                                        & \textbf{Thickness} & \textbf{Material}            \\
{Absorber}           & $\phi$ = 3.5 mm                                       & 40 $\mu$m          & diamond with bismuth coating \\
{Sensor (central)}   & 800$\times$250 $\mu$m$^{2}$                                            & 260 $\mu$m         & NTD germanium                \\
{Sensor (legs)}      & 3.4 mm & 150 $\mu$m         & NTD germanium                \\
{Mechanical Support} & 5.4 mm $\times$ 350 $\mu$m                                            & 68 $\mu$m          & sapphire                    
\end{tabular}
\caption{Dimensions of the bolometer.}
\label{tbl:Bolo184}
\end{table}

The detector is held 12.5 mm beneath the $\alpha$ particle source, which is collimated by a 2 mm diaphragm. The detector is orientated at a 45$^{\circ}$ angle relative to the source, inside an integrating sphere (part of its heritage mounting from the DIABOLO experiment~\cite{diabolo}, which coupled light from the feedhorn to the detector). The source, sphere, and detector are enclosed in a block to prevent stray light. The design specifications of the bolometer, as well as the experimental setup, have been described elsewhere~\cite{samalpha}\cite{samthese}\cite{setupDIABOLO}.\\

\begin{figure}[htbp]
    \vspace{0mm}
    \centering
    %\subfloat{{\includegraphics[width=.49\textwidth]{nonlint.png} }}%
    %\subfloat{{\includegraphics[width=.49\textwidth]{nonlinS.png} }}%
    \includegraphics[width=.6\textwidth]{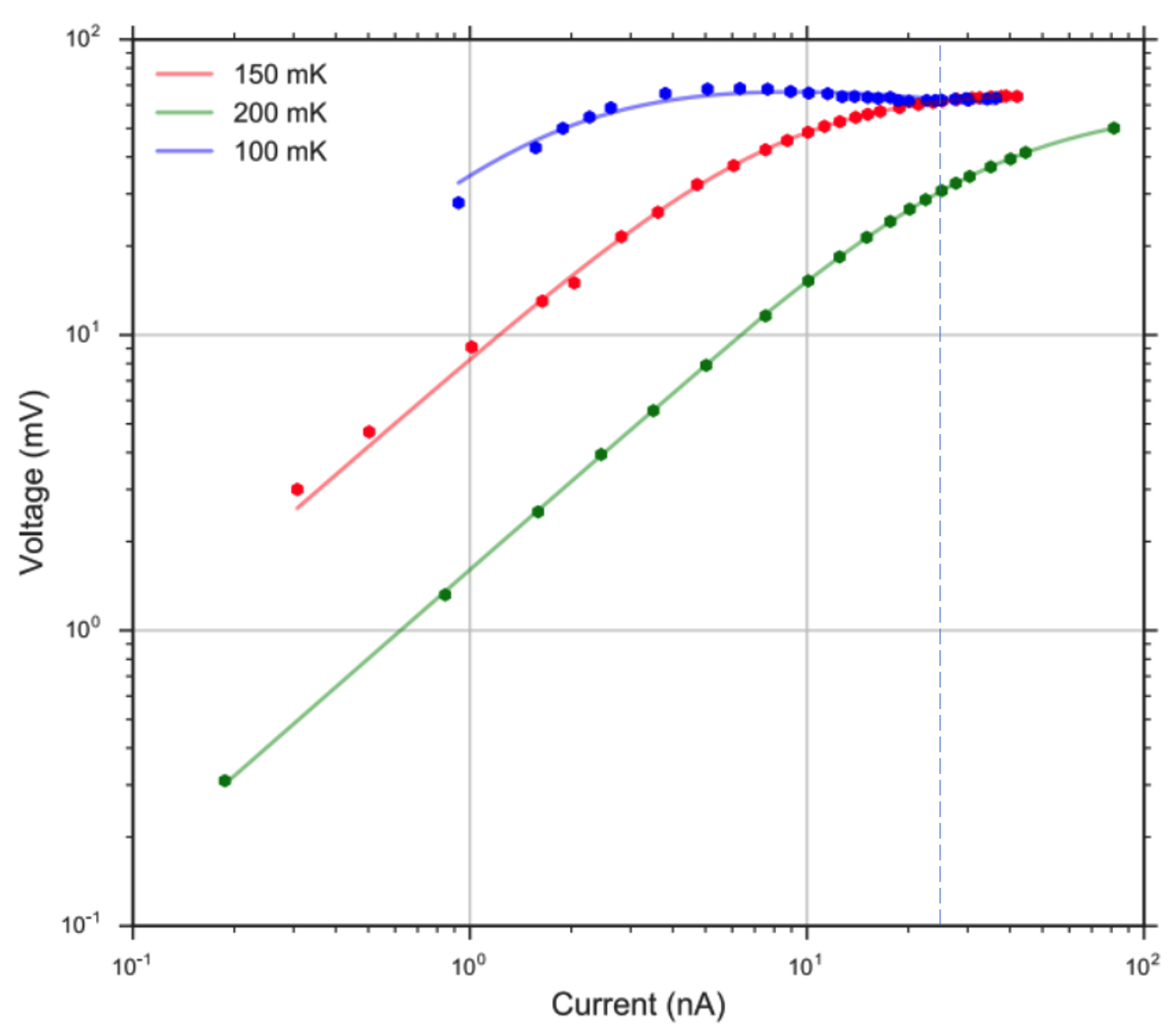}
    \caption{The load curve for Bolo 184 at 100 mK (blue), 150 mK (red), and 200 mK (green). Measured working points are shown as dots, and fitted load curves are shown as lines. The working point of the experiment described is shown by the blue line.}
    \label{fig:loadcurve}
    \vspace{-5mm}
\end{figure}
\vspace{0mm}

The bolometer was cooled in a dilution refrigerator to 100 mK and biased with $V_{\rm bias}$ = 1 V through a load resistor with $V_{\rm load} = 40 M \Omega$, corresponding to a load current of 25 nA. This load current is past the turnover point in the load curve, and thus past the optimal operation regime. The load curves for this bolometer, as well as the working point of this experiment, is shown in Fig.~\ref{fig:loadcurve}. At all shown temperatures, the measured working points are then fit to using synthetic load curves produced by the procedure outlined in Sudiwala et al. \cite{rashmi}.\\

\begin{figure}[htbp]
    \vspace{0mm}
    \centering
    %\subfloat{{\includegraphics[width=.49\textwidth]{nonlint.png} }}%
    %\subfloat{{\includegraphics[width=.49\textwidth]{nonlinS.png} }}%
    \includegraphics[width=.5\textwidth]{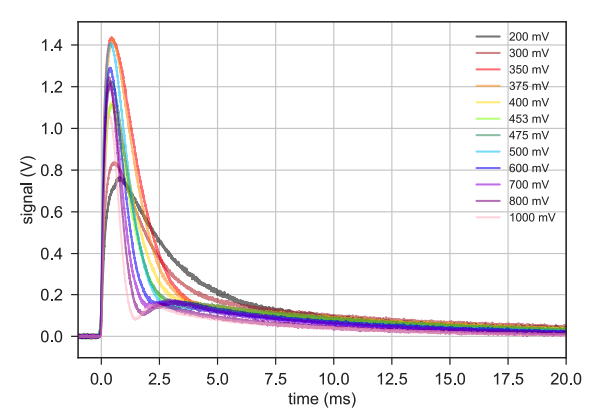}
    \caption{Average $\alpha$ particle response pulses in Bolo 184 with varying $V_{\rm bias}$}
    \label{fig:response}
    \vspace{-5mm}
\end{figure}
\vspace{0mm}

The transient response of the bolometer was recorded upon the absorption of $\alpha$ particles into the diamond absorber. At the high $V_{\rm bias}$ of 1 V, the impulse response to $\alpha$ particle energy deposition shows strong oscillatory features. These decrease as $V_{\rm bias}$ decreases, into which the bolometer is biased at a more stable point in its load curve. To demonstrate this, $V_{\rm bias}$ was varied between 200 mV and 1V. The average of all pulses taken at each $V_{\rm bias}$ illustrates this change in impulse response shape, and is shown in Fig.~\ref{fig:response}.\\

We suspect that this oscillatory behaviour is due to stray capacitance in the biasing circuit, as well as electro-thermal feedback in the bolometer. Confirmation of this effect is the subject of the modelling described in the next section.\\

\vspace{-7mm}
\section{Modelling}\label{sec:sims}
\vspace{-3mm}

\begin{figure}[htbp]
    \vspace{0mm}
    \centering
    %\subfloat{{\includegraphics[width=.49\textwidth]{nonlint.png} }}%
    %\subfloat{{\includegraphics[width=.49\textwidth]{nonlinS.png} }}%
    \includegraphics[width=.99\textwidth]{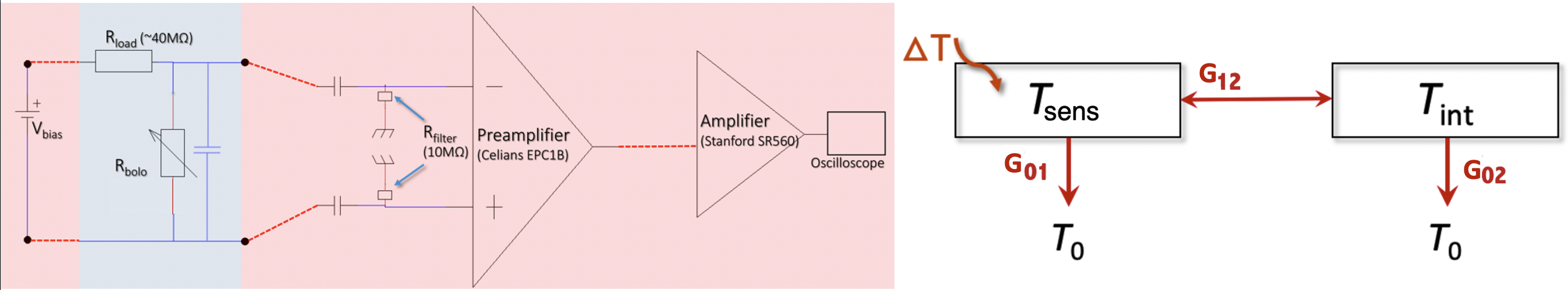}
    \caption{\textit{Left}: Electrical diagram of experimental setup. \textit{Right}: Block diagram of the bolometer thermal model.}
    \label{fig:blockmodel}
    \vspace{-5mm}
\end{figure}
\vspace{0mm}

Based on the thermistor design, the NTD sensor block is directly attached to the heatsink through a thermal conductance, and there is an intermediate stage which is also linked to the sensor and the thermal bath (analogous to the sapphire layer). A block diagram is shown in Fig.~\ref{fig:blockmodel} (right). The electrical diagram of the entire experimental setup is also shown in Fig.~\ref{fig:blockmodel} (left), with the effective stray capacitance of the cryogenic cables shown on the cold (blue) stage. \\

We find the starting bolometer working point from the $T_{0}$ (bath temperature, 100 mK), $V_{\textrm{sens}}$ (sensor voltage), $V_{\textrm{Bias}}$ (bias voltage), and $R_{\textrm{Load}}$ (resistance of the load resistor). We find the equilibrium point where $P_{\textrm{Joule}}$ = $P_{\textrm{Conduction}}$, which is described as:

\begin{equation}
\label{PJoule2}
P_{\textrm{Conduction}} = G_{01} \cdot (T_{\textrm{sens}}^{4}-T_{0}^{4}) + G_{12} \cdot (T_{\textrm{sens}} - T_{\textrm{int}})
\end{equation}

\noindent where $G_{12}$ is the thermal link between the intermediate stage and the sensor and $G_{01}$ is the thermal link between the sensor and the thermal bath. The $T^{4}$ dependence of $G_{01}$ comes from a $G$ with a $T^{3}$ conductance integrated along the link. Beginning with zero thermal power between the intermediate stage and the sensor ($P_{\textrm{int}}$):

\begin{equation}
\label{Pint2}
P_{\textrm{int}} = G_{12} \cdot (T_{\textrm{sens}} - T_{\textrm{int}}) - G_{02} \cdot (T_{\textrm{int}} - T_{0}) = 0
\end{equation}

The temperature of the intermediate stage is then found by:

\begin{equation}
\label{Pint3}
T_{\textrm{int}} = \frac{G_{12} \cdot T_{\textrm{sens}} + G_{02} \cdot T_{0}}{G_{12} + G_{02}}
\end{equation}

\noindent where $G_{02}$ is the thermal link between the intermediate stage and the thermal bath. This gives a Joule power of:

\begin{equation}
\label{PJoule3}
P_{\textrm{Joule}} = G_{01} \cdot  (T_{\textrm{sens}}^{4} - T_{0}^{4}) + \frac{G_{12} \cdot G_{02}} {G_{12} + G_{02}} \cdot (T_{\textrm{sens}} - T_{0})
\end{equation}

\noindent in which the second term arises from linear network laws.

An initial loop over temperature calculates $T_{\rm Sens}$, $R_{\textrm{sens}}$, $V_{\textrm{sens}}$, and $I_{\textrm{sens}}$ as a function of $T$. The Joule power on the sensor is found from $P$ = $V_{\textrm{sens}}^{2}$ / $R_{\textrm{sens}}$. %%The temperature of the intermediate stage is calculated from equation~\ref{Pint2}, and the total thermal link $G$ from:

%%\begin{equation}
%%\label{totalg}
%%G = g_{00} \cdot (T_{\textrm{sens}}^{4} - T_{0}^{4}) + G_{12} \cdot (T_{\textrm{sens}} - %%T_{\textrm{int}})
%%\end{equation}

The temperature of the intermediate stage is calculated from equation~\ref{Pint3}, and the total conduction from equation~\ref{PJoule2}. The loop continues until the total conduction is greater than the Joule power, stopping at the working point of the bolometer.\\

At the working point, a $\Delta T$ is injected into $T_{\textrm{sens}}$ and the new $R_{\textrm{sens}}$ is calculated. The system contains the same filters as those which exist in the amplifier readout chain of the experiment, with a $R_{\textrm{filter}}$ of 20 M$\Omega$, giving a time constant $\tau_{\textrm{filter}}$ = $R_{\textrm{filter}}$ $\cdot$ $C_{\textrm{stray}}$ where $C_{\textrm{stray}}$ is the stray capacitance on the output. We treat $C_{\rm stray}$ as an effective stray capacitance, in which the cryogenic cabling in the system is the dominant component. The typical signal rise time is $\tau$' = $R_{\textrm{Load}}$ $\cdot$ $C_{\textrm{stray}}$.\\
Looping over time, we take an additional time constant which accounts for all the resistances in the system:
\begin{equation}
\label{tauuuu}
\tau = \Big(\frac{1}{R_{\textrm{Load}}} + \frac{1}{R_{\textrm{sens}}} + \frac{1}{R_{\textrm{filter}}}\Big)^{-1} \cdot C_{\textrm{stray}}
\end{equation}
\noindent and the various time constants affect the signal via the following mechanism:
\begin{equation}
\label{tauuuuVB}
V_{\textrm{sens}}(t_{i}) - V_{\textrm{sens}}(t_{(i-1)}) = V_{\textrm{sens}}(t_{(i-1)}) + (\frac{V_{\textrm{Load}}}{\tau '} + \frac{V_{\textrm{Sens-WP}}}{\tau_{\textrm{filter}}} - \frac{V_{\textrm{sens}}(t_{(i-1)})}{\tau}) dt
\end{equation}
\noindent where $V_{\textrm{Sens-WP}}$ is the sensor voltage at the working point which is stored in memory before the $\Delta T$ injection. These and all other parameters are updated at each (small) $dt$. The signal is produced by subtracting $V_{\textrm{sens}}$ from $V_{\textrm{Sens-WP}}$. The final output of the model is a signal $V(t)$, which should allow for the reproduction of electro-thermal effects seen in the early double-peaked pulses, and how this behaviour changes at different temperatures. \\

\vspace{-7mm}
\subsection{Results}
In order to test the validity of the model, we compare its output with experimental data taken under the simulated conditions. As noted in prior studies, the pulse shape in this bolometer is partially dependent on the striking location of the $\alpha$ particle in the absorber, with impacts closest to the central thermistor producing the highest amplitude pulses~\cite{samalpha}. As we ignore the bolometer absorber in the thermal model, we expect better reproduction of the pulse shape when we compare it with the highest amplitude (most direct) pulses. In contrast, we expect that for smaller pulses, the absence of position dependency and the usual complexity of the bolometer would make it more difficult to reproduce the pulse shape using this simulation. To verify this, we choose a mid-amplitude pulse of 2 V, and check whether the simulation can reproduce its attributes.\\

\begin{figure}[htbp]
    \vspace{0mm}
    \centering
    %\subfloat{{\includegraphics[width=.49\textwidth]{nonlint.png} }}%
    %\subfloat{{\includegraphics[width=.49\textwidth]{nonlinS.png} }}%
    \includegraphics[width=.8\textwidth]{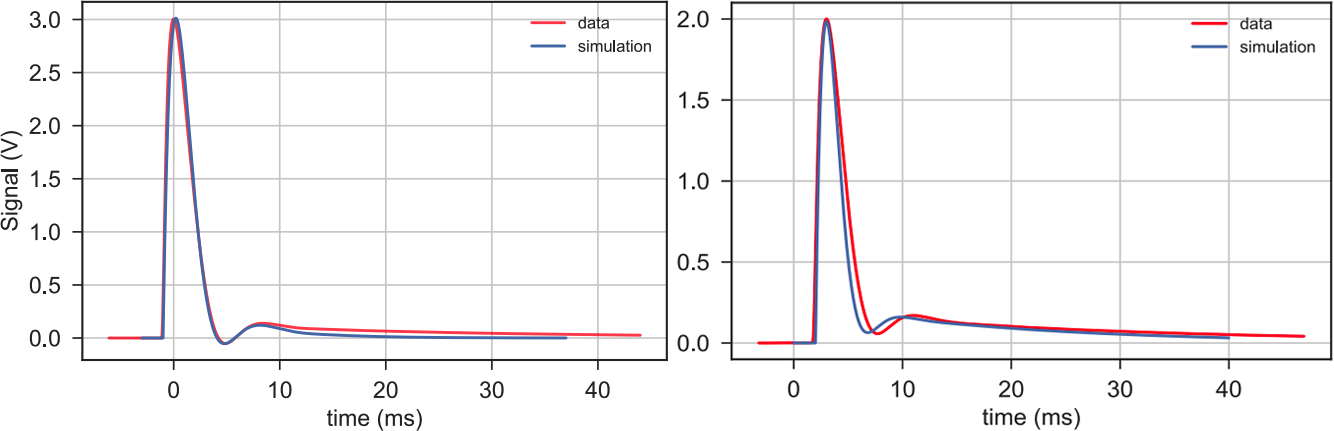}
\caption{Simulated pulses from the two-level thermal bolometer simulation (blue), with equal amplitude to a pulse from the 100 mK 1 V $V_{\textrm{bias}}$ dataset (red). Direct impact (left) is compared with the least direct (right).}
    \label{fig:2levelsim}
    \vspace{-5mm}
\end{figure}
\vspace{0mm}

We show a comparison between the measured and simulated pulses in Fig.~\ref{fig:2levelsim} (left). We find that the simulation quantitatively reproduces the shape of the largest-amplitude pulse, for $V_{\textrm{bias}}$ = 1 V, $R_{\textrm{Load}}$ = 40 M$\Omega$, $\Delta T$ = 0.014 K, $g_{00}$ = 320 \textrm{pW K$^{-1}$}, $C_{\textrm{stray}}$ = 0.15 nF, $G_{12}$ = \textrm{9.5 pW K$^{-1}$}, and  $G_{02}$ = \textrm{0.2 pW K$^{-1}$}. \\

Due to the complexity of balancing many free parameters, a $\chi^{2}$ minimisation routine was used to fit to these pulses, with known or measured parameters fixed. The lower-amplitude pulse, shown in Fig.~\ref{fig:2levelsim} (right), could not immediately be fit to using the parameters we have found above which produce the 3 V case. As we expect, the simulation is less able to reproduce the behaviour of a lower-amplitude pulse, where position-dependent effects from the absorber become significant \cite{samthese}. The shape remains qualitatively similar, but a small shift in time is observed due to the delayed heat flow from absorber to sensor in the data. Allowing for additional free parameters in the fitting routine, e.g. $R_{\textrm{Load}}$, achieved a higher $\chi^{2}$, but this lacks physical meaning.
Specifically, the best fit for this pulse is obtained using $R_{\textrm{Load}}$ of 65.89 M$\Omega$. This indicates that the thermal propagation effects are significant enough within the low-amplitude regime that one would have to produce a more complex simulation which accounts for the diamond absorber in order to produce the full pulse tail.\\
However, we have shown that the double-peaked behaviour in the high $V_{\textrm{bias}}$ 100 mK data set is most likely to be due to oscillations produced by the combination of electro-thermal coupling and stray capacitance on the bolometer electronics.\\

\vspace{-7mm}
\section{Conclusions}
\vspace{-3mm}
We have used a two-stage thermal model to simulate the effects of stray capacitance and electro-thermal feedback in the impulse response of a semiconductor bolometer operated at high $V_{\rm bias}$, with the goal of reproducing oscillatory features observed in laboratory measurements. The results of the model have been compared with experimental data obtained under the same conditions, where the bolometer response to $\alpha$ particle impacts has been measured. We find that the model qualitatively reproduces the pulse shape, having the best fitting results for high amplitude (central $\alpha$ hit) events, indicating a stray capacitance in the biasing circuit of 0.15 nF. However, the model does not take into account the bolometer absorber structure, and hence ignores the strongly position-dependent effects it produces. Due to this, the model does not fully account for the pulse shape when reproducing low-amplitude events, with a $\chi^{2}$ minimisation routine requiring non-physical parameters to produce a best fit. However, the simplified electrothermal model does qualitatively reproduce the pulse shapes obtained in the experimental data in absence of these position-dependent effects.\\

\vspace{-3mm}
\begin{acknowledgements}
\vspace{-3mm}
This work was supported by World Premier International Research Center Initiative (WPI), MEXT, Japan.
\end{acknowledgements}
\vspace{-3mm}
\vspace{-3mm}
\section*{Declarations}
\vspace{-3mm}
The authors declare that they have no conflict of interest. The datasets generated during and/or analysed during the current study are not publicly available due to archival but are available from the corresponding author on reasonable request.

\end{document}